\documentclass[twocolumn,preprintnumbers,twoside,slac_two]{revtex4}
\usepackage{graphicx}
\usepackage{fancyhdr}

  \newcommand {\be}{\begin{equation}}
\newcommand {\ee}{\end{equation}}
\begin{document}

\title{On the Sources of CP-violation Contributing to the Electric Dipole Moments
}

%

\author{Durmu{\c s} A. Demir }
\affiliation{Department of Physics, Izmir Institute of Technology,
Izmir, TR 35430, Turkey}
\author{Yasaman Farzan}
\affiliation{Institute for Studies in Theoretical Physics and
Mathematics (IPM), PO Box 19395-5531, Tehran, Iran}

\begin{abstract}
In the framework of seesaw mechanism embedded in the constrained
Minimal Supersymmetric Standard Model (cMSSM), phases of neutrino
Yukawa coupling, $\mu$-term and $A$-terms can all contribute to
the Electric Dipole Moment (EDM) of the electron. We discuss and
classify the situations for which by combined analysis of the
upcoming results on $d_e$, $d_{{\rm Hg}}$ and $d_D$ discriminating
between these sources will be possible.
\end{abstract}

\maketitle

\thispagestyle{fancy}


\section{Introduction}
Recent atmospheric and solar neutrino data \cite{solar} as well as
the results of  KamLAND \cite{kamland}, K2K \cite{k2k} and MINOS
\cite{minos} establish nonzero mass for neutrinos. On the other
hand, kinematical studies \cite{mainz} and cosmological
observations \cite{cosmo} show that neutrino masses cannot exceed
a few eV. Nonzero neutrino masses cannot be accommodated within
the Standard Model (SM). Several models have been suggested to
attribute  tiny yet nonzero masses for neutrinos among which the
seesaw mechanism \cite{seesaw} is arguably the most popular one.
The seesaw mechanism adds three SM singlet right-handed neutrinos
with very heavy Majorana masses to the model. Up to now, we have
no clue how large the masses of the right-handed neutrinos are.
The scale can lay anywhere between TeV up to $10^{14}$~GeV. If the
scale is higher than 1 TeV, the right-handed neutrinos cannot be
produced by accelerator technology; thus, we can only learn about
the seesaw parameters indirectly through their effects on low
energy parameters \cite{sacha}, such as light neutrino mass matrix
and slepton mass matrix in the context of supersymmetric seesaw
\cite{franchesca}. In this line, any  low energy observable which
is sensitive to the seesaw parameters deserves special attention.

The $3\times 3$ neutrino Yukawa matrix  introduces six new sources
of CP-violation which, in the context of supersymmetric standard
model, can induce significantly large contributions to the
electric dipole moments of the electron
\cite{YukawaEDM,manopeskin} and other particles. So far no finite
EDM has been observed \cite{pdg}:
\begin{equation} |d_e|<1.4\times 10^{-27}\ \mbox{e~cm}.\end{equation}
However, there are ongoing experiments \cite{ongoing} as well as
proposals \cite{proposals} to improve the present bound by several
orders of magnitude. In view of these experiments, it has been
suggested to use the EDMs to extract information on the seesaw
parameters \cite{dutta}. However, for deriving any information
from $d_e$ we must be aware of other sources of CP-violation that
can give a significant contribution to $d_e$.

In the cMSSM, there are two extra sources of CP-violation relevant
for the EDMs of leptons: the  phases of the $\mu$ parameter and
the universal trilinear coupling $a_0$.
 The phases of $a_0$ and $\mu$  induce $d_e\sim (m_e/m_d) d_d\sim (m_e/m_u)d_u
\sim e(m_e/m_d) \tilde{d}_d\sim e (m_e/m_u)\tilde{d}_u$, where
$\tilde{d}_u$ and $\tilde{d}_{d}$,  respectively, are the Chromo
EDMs (CEDM) of up and down quarks which give rise to EDMs of
hadrons and nuclei such as mercury  and deuteron. However, the
quark EDMs and CEDMs induced by the phases of $Y_\nu$ are too
small to be detectable in near future. Therefore, if complex
$Y_\nu$ is the only source of CP-violation, we expect the Deuteron
EDM ($d_D$) to be too small to be detectable in the near future
($d_D$ is measured with ionized deuteron which is depleted from
electrons). Based on this difference, it has been suggested in
\cite{durmush} to combine the information on $d_D$ and $d_{{\rm
Hg}}$ with $d_e$ to disentangle the source of CP-violation. It is
also discussed how much the present bounds have to be improved in
order to be able to make such a discrimination. In present letter,
we review the results obtained in \cite{durmush}.

\section{The model}
The seesaw mechanism embedded in the MSSM is described by the
superpotential
 \be
W=Y_\ell^{i j}E_{i} H_d\cdot  L_j- Y_\nu^{i j}N_{i} H_u \cdot
L_j+\frac{1}{2}M_{ij} N_{i} N_{j}-\mu  H_d \cdot H_u . \ee The
quark Yukawa couplings, not shown here, are the same as in the
MSSM. Here, $i,j$ are generation indices, $L_{j }$ consist of
lepton doublets $(\nu_{j L}, \ell^-_{jL})$, and $E_i$ contain
left-handed anti-leptons $\ell^+_{iL}$. The superfields $N_i$
contain anti right-handed neutrinos. Without loss of generality,
one can rotate and rephase the fields to make Yukawa couplings of
charged leptons ($Y_\ell$) as well as the mass matrix of the
right-handed neutrinos ($M_{ij}$)  real diagonal. In what follows,
we will use this basis.

In general, the soft supersymmetry breaking terms (the
mass-squared matrices and trilinear couplings of the sfermions)
can possess flavor-changing entries which facilitate a number of
flavor-changing neutral current processes in the hadron and lepton
sectors. The existing experimental data thus put stringent bounds
on flavor-changing entries of the soft terms. For instance,
flavor-changing entries of the soft terms in the lepton sector can
result in sizeable $\mu \to e \gamma$, $\tau \to e \gamma$ and
$\tau \to \mu \gamma$.
This motivates us to go to the  constrained MSSM framework where
soft terms of a given type unify at a scale close to the scale of
gauge coupling unification. In other words, at the GUT scale, we
take
\begin{eqnarray} {\cal L}_{soft} &=&
-m_0^2(\tilde{L}_{i}^\dagger\tilde{L}_{i}+ \tilde{E}_{i}^\dagger
\tilde{E}_{i}+ \tilde{N}_{i}^\dagger \tilde{N}_{i}+H_d^\dagger
H_d+ H_u^\dagger H_u \label{soft}) \cr &-&
 \frac{1}{2} m_{1/2}(\tilde{B} \tilde{B}+
\tilde{W} \tilde{W}+\tilde{g} \tilde{g}+{\rm H.c.}) \cr &-& (
\frac{1}{2} B_H H_d \cdot H_u +{\rm H.c.})\cr &-&(
A_\ell^{ij}\tilde{E}_{i} H_d\cdot \tilde{L}_{j}-
A_\nu^{ij}\tilde{N}_{i} H_u\cdot  \tilde{L}_{j} +{\rm H.c.})\cr
&-& (\frac{1}{2} B_\nu M_i \tilde{N}^i\tilde{N}^i+{\rm H.
c.}).\label{soft} \end{eqnarray} Here $A_\ell=a_0 Y_\ell$ and
$A_\nu=a_0 Y_\nu$, where $a_0$ in general can be complex and a
source of CP-violation. The last term is the lepton number
violating neutrino bilinear soft term which is called the neutrino
$B$-term. As has been first shown in \cite{yasaman}, the phase of
the neutrino $B$-term can induce a contribution to $d_e$. In this
letter, for simplicity, we will set $B_\nu=0$. By rephasing the
Higgs fields we can make $B_H$ real; however, the phase of $\mu$
will in general remain nonzero. In this letter, to calculate the
effects of the phases of $Y_\nu$, $\mu$ and $a_0$ on the EDMs and
CEDMs, we use the results of \cite{durmush,manopeskin,nath}.
\section{Bounds on the EDMs} In this section, we first review the current bounds on $d_D$, $d_{Hg}$ and $d_n$ and the prospects for improving them. We
then review how we can write them in terms of Im$(\mu)$ and
Im$(a_0)$.
\begin{itemize}

\item {\bf Neutron EDM, $d_n$:} The present bound on
$d_n$ \cite{dn} is \be d_n<3.0\times 10^{-26} e~{\rm cm} \ \ {\rm
at }\ \ 90 \% \ \ {\rm C.L.}\ee This bound can be improved
considerably by SNS \cite{lan} which will be able to probe $d_n$
down to $ 10^{-28}~e$~cm.
\item {\bf Mercury EDM, $d_{Hg}$:}
The present bound on $d_{Hg}$ is \be \label{mercury} |d_{Hg}|<2.1
\times 10^{-28} e~{\rm cm} \ee which can be improved by a factor
of 2 \cite{hgexp}.

\item {\bf Deuteron EDM, $d_D$:}  The present bound on $d_D$ is too weak to constrain
the CP-violating phases; however, there are proposals \cite{Dexp}
to probe $d_D$ down to \be (1-3)\times 10^{-27}~ e~{\rm cm}
\label{semer}. \ee
\end{itemize}Different sources of
CP-violation affect the EDMs listed above differently. As a
result, in principle, by combining information on these
observables we can discriminate between different sources of
CP-violation. It is rather straightforward to calculate the EDMs
and CEDMs of the elementary particles in terms of the phases of
$a_0$ and $\mu$ \cite{durmush,manopeskin,nath}; however, writing
$d_n$, $d_{Hg}$ and $d_D$ in terms of  the EDMs and CEDMs of their
constituents is more difficult and a subject of debate among the
experts.  Let us consider them one by one.
\begin{itemize}
\item {\bf $d_n(d_q,\tilde{d}_q)$:}
\newline
Despite the  rich literature on $d_n$ in terms of the quark EDMs
and CEDMs, the results are quite model dependent. For example, the
SU(3) chiral model \cite{su(3)chiral} and QCD sum rules
\cite{sum-rules} predict different contributions from
$\tilde{d}_u$ and $\tilde{d}_d$ to $d_n$. Considering these
discrepancies in the literature,  we do not use bounds on $d_n$ in
our analysis.
\item
{\bf $d_{Hg}(d_q,\tilde{d}_q)$:}
\newline
There is an extensive literature on $d_{Hg}$ \cite{dhg}. Following
Ref. \cite{210-26}, we will interpret the bound on $d_{Hg}$ as \be
|\tilde{d}_d-\tilde{d}_u|<2 \times 10^{-26}~ {\rm cm}.\ee
\item {\bf $d_D(d_q,\tilde{d}_q)$:}
\newline
Searches for  $d_D$ can serve as an ideal probe for the  existence
of sources of CP-violation other than complex $Y_\nu$ because $i)$
there is a good prospect of improving the bound on $d_D$
\cite{Dexp}; $ii)$  an ionized deuteron does not contain any
electrons and hence we expect only a negligible and undetectable
contribution from $Y_\nu$ to $d_D$.

To calculate $d_D$ in terms of quark EDMs and CEDMs, two
techniques have been suggested in the literature: $i)$ QCD sum
rules \cite{lebedev} and $ii)$  SU(3) chiral theory \cite{dsu(3)}.
Within the error bars, the two models agree on the contribution
from $\tilde{d}_d-\tilde{d}_u$ which is the dominant one. However,
the results of the two models on the sub-dominant contributions
are not compatible. Apart from this discrepancy, there is a large
uncertainty in the contribution of the dominant term: \be
d_D(d_q,\tilde{d}_q)\simeq
-e(\tilde{d}_u-\tilde{d}_d)\,5^{+11}_{-3} \ . \ee In this paper we
take ``the best fit" for our analysis.

\end{itemize}
\section{Numerical analysis}
\begin{figure}
\includegraphics[width=80 mm,bb=50 28 545 477, clip=true]{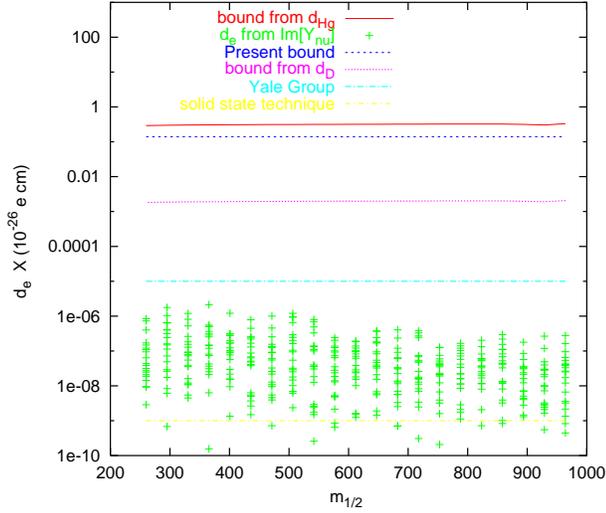} \caption{ Electron EDM for $a_0=0$, $\tan \beta=10$ and
sgn($\mu)=+$. The scatter points represent $|d_e|$ induced by
random complex $Y_\nu$ compatible with the data. The blue dashed
line is the present bound on $d_e$ \cite{pdg} and dot-dashed lines
show the values of $d_e$ that can be probed in the future
\cite{ongoing,proposals}}
 \label{a00tan10}
\end{figure}

\begin{figure}
\includegraphics[width=80 mm,bb=50 28 545 477, clip=true]{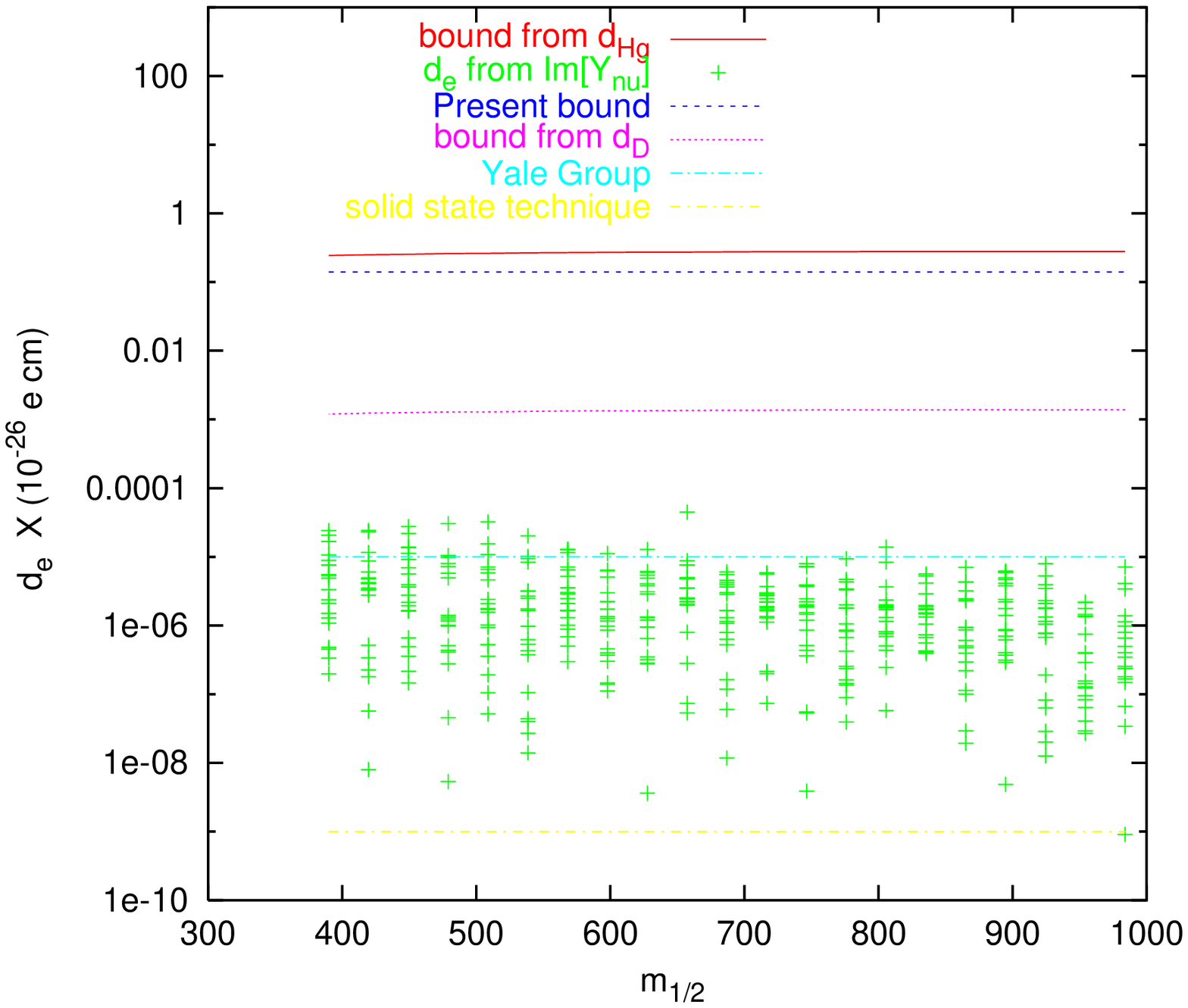}
\caption{Electric dipole moment of the electron for
$a_0=1000$~GeV, $\tan \beta=10$ and sgn($\mu)=+$. To draw the red
solid and purple dotted lines, we have assumed that Im[$a_0$] is
the only source of CP-violation and have taken
$\tilde{d}_d-\tilde{d}_u$ equal to  $2 \times 10^{-26}$~cm and $2
\times 10^{-28}$~cm, respectively to derive  Im[$a_0$]. To produce
the scatter points, we have assumed that complex $Y_\nu$ is the
only source of CP-violation and have randomly produced $Y_\nu$
compatible with the data. The blue dashed line is the present
bound on $d_e$ \cite{pdg} and  dot-dashed lines show the values of
$d_e$ that can be probed in the future \cite{ongoing,proposals}}
 \label{a01000tan10}
\end{figure}

\begin{figure}
\includegraphics[width=80 mm,bb=50 28 545 477, clip=true]{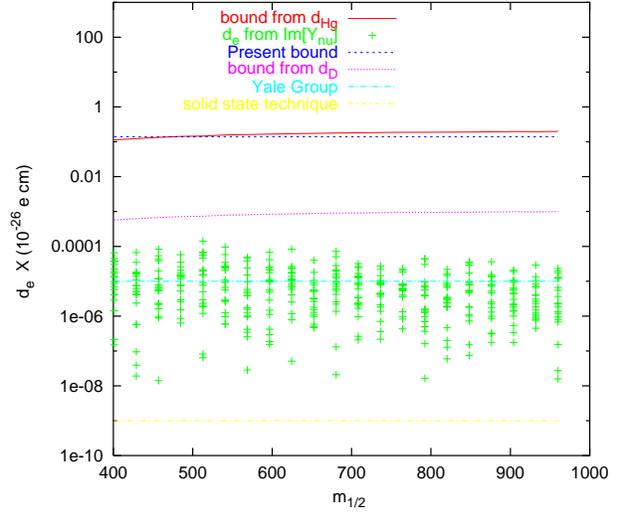}
\caption{The same as fig. \ref{a01000tan10} for $a_0=1000$~GeV,
$\tan \beta=20$ and sgn($\mu)=+$.} \label{a01000tan20}
\end{figure}

\begin{figure}
\includegraphics[width=80 mm,bb=50 28 545 477, clip=true]{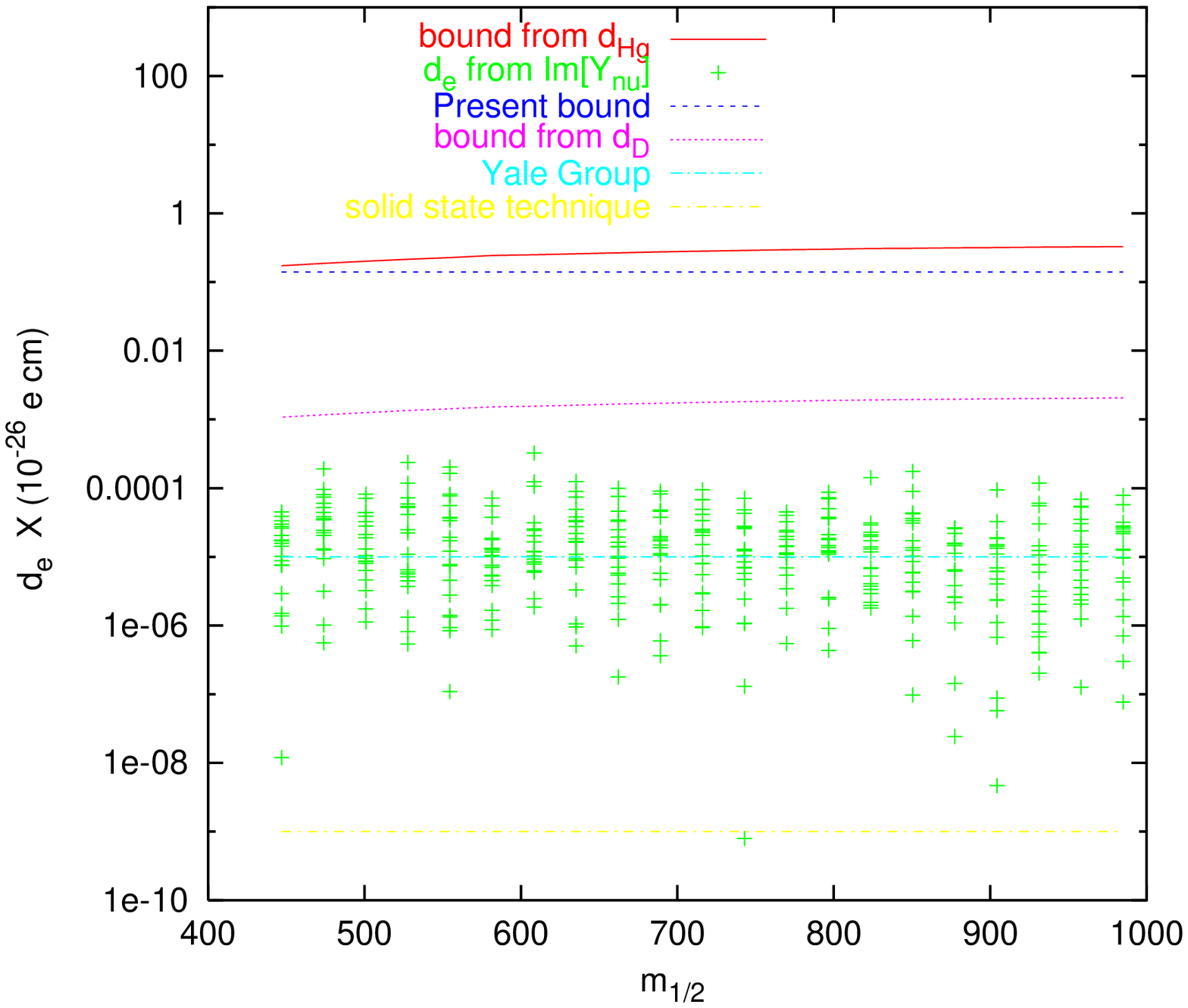} \caption{Electric dipole moment of the
electron for $a_0=2000$~GeV, $\tan \beta=20$ and sgn($\mu)=+$. To
draw the red solid and purple dotted lines, we have assumed that
Im[$\mu$] is the only source of CP-violation and have taken
$\tilde{d}_d-\tilde{d}_u$ equal to  $2 \times 10^{-26}$~cm and $2
\times 10^{-28}$~cm, respectively to derive Im[$\mu$]. To produce
the scatter points, we have assumed that complex $Y_\nu$ is the
only source of CP-violation and have randomly produced $Y_\nu$
compatible with the data. The blue dashed line is the present
bound on $d_e$ \cite{pdg} and  dot-dashed lines show the values of
$d_e$ that can be probed in the future \cite{ongoing,proposals}}
 \label{a02000tan20}
\end{figure}

In this section, we first describe how we  produce the random
seesaw parameters compatible with the data. We then describe
figs. (\ref{a00tan10}-\ref{a02000tan20}) and, in the end, discuss
what can be inferred from the future data considering different
possible situations one by one.

In figures (\ref{a00tan10}-\ref{a02000tan20}), the scatter points
marked with "+" represent $d_e$ resulting from complex $Y_\nu$. To
extract random $Y_\nu$ and $M_N$ compatible with data, we have
followed the recipe described in \cite{recipe} and solved the
following two equations \be  Y_\nu^T {1\over M}Y_\nu (v^2\sin^2
\beta) /2=U\cdot \Phi \cdot M_\nu^{Diag} \cdot \Phi \cdot U^T \ee
and
\begin{eqnarray} h\equiv Y_\nu^\dagger {\rm Log}{ M_{GUT} \over M}
Y_\nu=\left[ \matrix{ a &0 & d \cr 0& b& 0 \cr d^* & 0& c}\right],
\label{h}\end{eqnarray} where $v=247$ GeV, $M$ is the mass matrix
of the right-handed neutrinos, $U$ is the mixing matrix of
neutrinos with $s_{13}=0$ and $\Phi$ is $diag[1,e^{i\phi_1},e^{i
\phi_2}]$ with random values of $\phi_1$ and $\phi_2$ in the range
$(0,2\pi)$. Finally, $M_\nu^{Diag}= diag[m_1,\sqrt{m_1^2+\Delta
m_{21}^2},\sqrt{m_1^2+\Delta m_{31}^2}]$ where $m_1$ picks up
random values between 0 and 0.5 eV in a linear scale. The upper
limit on $m_1$   is what has been found in \cite{hannestad} by
taking the dark energy equation of state a free (but constant)
parameter.

 In order to satisfy the strong bounds on
$\mbox{Br}(\mu \to e \gamma$) \cite{pdg} and $\mbox{Br}(\tau \to
\mu \gamma )$ \cite{newboundontaumu}, the matrix $h$, defined in
Eq. (\ref{h}), is taken to have this specific pattern with zero $e
\mu$ and $\mu \tau$ elements. Actually these branching ratios put
bounds on $(\Delta m_{\tilde{L}}^2)_{e \mu}$ and $(\Delta
m_{\tilde{L}}^2)_{\mu \tau}$ rather than on $h_{e \mu}$ and
$h_{\mu \tau}$. Notice that only the dominant term of $\Delta
m_{\tilde{L}}^2$ is proportional to $h$. There is also a
subdominant ``finite" contribution to $\Delta m_{\tilde{L}}^2$
which is about 10\% of the dominant effect and is not proportional
to the matrix $h$ \cite{manopeskin}. Nonetheless, for extracting
the seesaw parameters, 20\% accuracy is enough and we can neglect
the subdominant part and take $\Delta m_{\tilde{L}}^2$
proportional to the matrix $h$. In Eq. (\ref{h}), $a$, $b$, $c$
are real numbers which take random
 values
  between 0 and 5. On the other hand, $|d|$ takes random values between 0 and the
 upper bound from $\mbox{Br}(\tau \to e \gamma)$ \cite{newboundontaue}. To calculate
 the upper  bound on $|d|$, we have used the formulae derived in
Ref. \cite{carlos}. The phase of $d$  takes random values between
0 and 2$\pi$.

To perform this analysis we have taken various values of $\tan
\beta$ and $a_0$ and calculated the spectrum of the supersymmetric
parameters along the $m_{1/2}-m_0$ strips parameterized in Ref.
\cite{petra}. Notice that Ref. \cite{petra} has already removed
the parameter range for which color or charge condensation takes
place.

In the figures, we have also drawn the present bound on $d_e$
\cite{pdg} as well as the limits  which can be probed in the
future. The present bound is shown by a dashed dark blue line and
lies several orders of magnitude above the $d_e$ from phases of
$Y_\nu$. After five years of data-taking, the Yale group can probe
$d_e$ down to $10^{-31}$ {\it e}~cm \cite{ongoing} which is shown
with a dot-dashed cyan line in the figures. As  demonstrated in
the figures, only  for large values of $a_0$ the effect of complex
$Y_\nu$ on $d_e$ can be probed by the Yale group and for most of
the parameter space the effect remains beyond the reach of this
experiment.

There are proposals \cite{proposals} to use solid state techniques
to probe $d_e$ down to $10^{-35}$ {\it e}~cm (shown with
dot-dashed yellow line in the figure). In this case, as it can be
deduced from the figure, we will have a great chance of being
sensitive to the effects of the phases of $Y_\nu$ on $d_e$.
However, unfortunately, the feasibility and time scale of the
solid state technique is still uncertain.

In figs. \ref{a00tan10} and \ref{a02000tan20}, $d_e$ resulting
from Im[$\mu$] is also depicted. The red solid lines in these
figures show $d_e$ from Im[$\mu$] assuming that the corresponding
$d_{Hg}$ saturates the present bound  \cite{hgexp}. As  is
well-known, there are uncertainties both in the value of $m_d$
\cite{pdg} and in the interpretation of $d_{Hg}$ in terms of more
fundamental parameters $\tilde{d}_u$, $\tilde{d}_d$ and
$\tilde{d}_s$. To draw this curve we have assumed $m_d=5$ MeV and
$\tilde{d}_u-\tilde{d}_d<2 \times 10^{-26}$ {\it e}~cm . As shown
in the figure this bound is weaker than even the present direct
bound on $d_e$.
 The purple dotted lines in figs. (\ref{a00tan10}, \ref{a02000tan20}),
 represent $d_e$ induced by values of Im$[\mu]$ that give rise to
$\tilde{d}_u-\tilde{d}_d=2\times 10^{-28}$ cm (corresponding to
$d_{D}=10^{-27}$ {\it e}~cm  and $d_D= 5e(\tilde{d}_d-\tilde{d}_u)
$). Notice that these  curves lie  well below the direct bound on
$d_e$ but the Yale group will be able to probe even smaller values
of $d_e$. Similarly in figs.
(\ref{a01000tan10},\ref{a01000tan20}), $d_e$ resulting from
Im[$a_0$] is depicted.

 In the figures, the bounds from $d_{Hg}$ and $d_D$ appear
 almost as horizontal lines. This results from the fact that for the $m_0-m_{1/2}$
 strips that
 we analyze, $m_0$ is almost proportional to $m_{1/2}$. Using dimensional analysis
 we can write
 $$ \tilde{d}_u-\tilde{d}_d \simeq k_1 {{\rm Im}[\mu] \ \ {\rm or \ \ Im}[a_0] \over
 m_{1/2}^3} \ \ \ \ \ \ d_e \simeq k_2 {{\rm Im}[\mu]  \ \ {\rm or \ \ Im}[a_0]\over m_{1/2}^3 }$$
 where $k_1$ and $k_2$ are given by the relevant fermion masses
 and are independent of $m_{1/2}$.
 As a result, for a given value of $\tilde{d}_u-\tilde{d}_d $,
 Im[$\mu$] (or Im[$a_0$]) itself is proportional to $m_{1/2}^3$ so
 $d_e$ will not vary with $m_{1/2}$.

In the following, we will discuss what can be inferred about the
sources of CP-violation from $d_e$ and $d_D$ if their values (or
the bounds on them) turn out to be in certain ranges.
According to  Fig.~\ref{a00tan10}, for $a_0=0$, any signal found
by the Yale group implies that there are sources of CP-violation
other than the phases of the Yukawa couplings. However, for larger
values of $a_0$, the effect of $Y_\nu$ on the EDMs can be observed
by the Yale group within five years. According to Figs.
(\ref{a01000tan10}-\ref{a02000tan20}), for $a_0 \stackrel{>}{\sim}
1000$~GeV  EDMs originating from complex $Y_\nu$ can be large
enough to be observed by the Yale group. Therefore, if after five
years the Yale group reports a null result, we can derive bounds
on certain combinations of seesaw parameters and $a_0$. At least
it will be possible to discriminate between low and high $a_0$
values.  However, if the Yale group finds that $ 10^{-31}~e~{\rm
cm}<d_e<10^{-29}$~{\it e}~cm we will not be able to determine
whether $d_e$ originates from complex $Y_\nu$ or from more
familiar sources such as complex $a_0$ or $\mu$. To be able to
make such a distinction, values of $d_D$ down to
$10^{-28}-10^{-29}$ {\it e}~cm   have to be probed which, at the
moment, does not seem to be achievable.

 If future searches for $d_D$
find  $d_D>10^{-27}$ {\it e}~cm  but the Yale group finds $d_e<2
\times 10^{-29}$ {\it e}~cm  (this can be tested within only 3
years of data taking by the Yale group \cite{ongoing}), we might
conclude that the source of CP-violation is something other than
pure Im$[\mu]$ or Im$[a_0]$; {\it e.g.,} the QCD $\theta$-term
which can give a significant contribution to $d_D$ but only a
negligible contribution to $d_e$. Another possibility is that
there is a cancelation between the contributions of Im$[\mu]$ and
Im$[a_0]$ to $d_e$ \cite{cancelation}. The information on $d_n$
would then help us to resolve this ambiguity provided that the
theoretical uncertainties in calculation of $d_n$ as well  as
$d_D$ are sufficiently reduced.

On the other hand, if the Yale group detects $d_e>2\times
10^{-29}$ {\it e}~cm, we will expect that $d_D>10^{-27}$ {\it
e}~cm  which will be a strong motivation for building a deuteron
storage ring and searching for $d_D$. If such a detector finds a
null result, within this framework the explanation will be quite
non-trivial requiring some fine-tuned cancelation between
different contributions.

According to these figures, in the foreseeable future, we will not
be able to extract any information on the seesaw parameters from
EDMs, because even if we develop techniques to probe $d_e$ as
small as $10^{-35}\, e~ {\rm cm}$, we will not be able to subtract
(or dismiss) the effect coming from Im$[\mu]$ and Im$[a_0]$ unless
we are able to probe $\tilde{d}_u-\tilde{d}_d$ at least 5 orders
of magnitude below its present bound which seems impractical.
Remember  that this is under the optimistic assumptions that the
mass of the lightest neutrino, $m_1$, and $\mbox{Br}(\tau \to e
\gamma)$ are close to their upper bounds and there is no
cancelation between different contributions to the EDMs.

  If, in the future, we realize that $m_1$ and $\mbox{Br}(\tau
\to e \gamma)$ are indeed close to the present upper bounds on
them and $a_0=0$ ($a_0=1000\ {\rm GeV}$)  but find
$d_e<10^{-35}${\it e}~cm ($d_e<10^{-34}$ {\it e}~cm ), we will be
able to draw bounds on the phases of $Y_\nu$ which along with the
information on  the Dirac and Majorana phases of the neutrino mass
matrix and the CP-violating phase of the left-handed slepton mass
matrix may have some implication for leptogenesis. This is however
quite an unlikely situation.

\section{Summary}
In this work we have studied EDMs of particles in the context of
supersymmetric seesaw mechanism.
 In
figs. \ref{a00tan10}-\ref{a02000tan20}, the values of $d_e$
corresponding to different random complex $Y_\nu$ textures are
represented by ``+". For small values of $\tan \beta$ ($\tan
\beta<10$) and $a_0$ ($a_0<1000$~{\rm GeV}), $d_e$ induced by
$Y_\nu$ is beyond the reach of the ongoing experiments
\cite{ongoing}. Such values of $d_e$  can however be probed by the
proposed solid state  based  experiments \cite{proposals}. For
larger values of $\tan \beta$ and/or $a_0$, the Yale group may be
able to detect the effects of complex $Y_\nu$ on $d_e$. As
demonstrated in Figs. \ref{a01000tan20} and \ref{a02000tan20}, for
$\tan \beta=20 $ and $a_0=1000-2000$ GeV, a large fraction of
parameter space yields $d_e$  detectable by the Yale group.
However, even in this case we will not be able to extract
information on the seesaw parameters from $d_e$ because the source
of CP-violation might be $a_0$ and/or $\mu$ rather than $Y_\nu$.
If the future searches for $d_D$ \cite{lebedev} find out that
$d_D>10^{-27}$ {\it e}~cm  then we will conclude that there is a
source of CP-violation other than complex $Y_\nu$. However, to
prove that the dominant contribution to $d_e$ detected by the Yale
group comes from complex $Y_\nu$-- hence to be able to extract
information on the seesaw parameters from it-- we should show that
$d_D<10^{-28}-10^{-29}$ {\it e}~cm which is beyond the reach of
even the current proposals. Notice that for the purpose of
discriminating between complex  $Y_\nu$ and $a_0/\mu$ as sources
of CP-violation, searching for $d_{Hg}$ is not very helpful
because mercury atom contains electron and hence $d_{Hg}$ obtains
a contribution from complex $Y_\nu$. That is while ionized
deuteron used for measuring $d_D$ does not contain any electron
and the contribution of complex $Y_\nu$ to it is negligible. To
obtain information from $d_n$, the theoretical uncertainties first
have to be resolved.

\begin{acknowledgments}
D.A.D. is grateful to Institute for Studies in Theoretical Physics
and Mathematics (IPM) for its generous hospitality while the work
on which this talk is based was prepared, and International Centre
for Theoretical Physics (ICTP), Turkish Academy of Sciences
(through GEBIP grant), and Scientific and Technical Research
Council of Turkey (through project 104T503) for their financial
supports.
\end{acknowledgments}

\bigskip 

\end{document}